\documentclass[aps,prb,showpacs,twocolumn,superscriptaddress,preprintnumbers,amsmath,amssymb]{revtex4}
\usepackage{graphicx}
\begin{document}
\title{Comparative study of helimagnets MnSi and Cu$_2$OSeO$_3$ at high pressures}

\affiliation{Institute for Solid State Physics, Russian Academy of Sciences, Chernogolovka, Moscow Region, Russia}
\author{V.A. Sidorov}
\affiliation{Institute for High Pressure Physics of Russian
Academy of Sciences, Troitsk, Russia}
\author{A.E. Petrova}
\affiliation{Institute for High Pressure Physics of Russian
Academy of Sciences, Troitsk, Russia}
\author{P.S. Berdonosov}
\affiliation{Department of Chemistry, Moscow State University, Moscow, Russia}
\author{V.A. Dolgikh}
\affiliation{Department of Chemistry, Moscow State University, Moscow, Russia}
\author{S.M. Stishov}
\email{sergei@hppi.troitsk.ru}
\affiliation{Institute for High Pressure Physics of Russian Academy of Sciences, Troitsk, Russia}

\begin{abstract}
The heat capacity of helical magnets Cu$_2$OSeO$_3$ and MnSi has been investigated at high pressures by the ac-calorimetric technique. Despite the differing nature of their magnetic moments, Cu$_2$OSeO$_3$ and MnSi demonstrate a surprising similarity in behavior of their magnetic and thermodynamic properties at the phase transition. Two characteristic features of the heat capacity at the phase transitions of both substances (peak and shoulder) behave also in a similar way at high pressures if analyzed as a function of temperature. This probably implies that the longitudinal spin fluctuations typical of weak itinerant magnets like MnSi contribute little to the phase transition.
The shoulders of the heat capacity curves shrink with decreasing temperature suggesting that they arise from classical fluctuations. In case of MnSi the sharp peak and shoulder at the heat capacity disappear simultaneously probably signifying the existence of a tricritical point and confirming the fluctuation nature of the first order phase transition in MnSi as well as in Cu$_2$OSeO$_3$.
\end{abstract}
\maketitle
\section{Introduction}
Manganese silicide (MnSi), a model itinerant helimagnet , crystallizes in a B20 structure, whose non-centro-symmetric space group P2$_1$3 allows a helical (chiral) magnetic structure. The phase transition in MnSi from helical to paramagnetic states at $\approx$30 reveals some remarkable features such as sharp peaks and shoulders on the high temperature side of the peaks in the heat capacity, thermal expansion, temperature dependence of resistivity and sound absorption~\cite{1,2,3,4,5,6,7}. As is recognized now the sharp peaks in the above properties indicate a first order nature of the phase transition whereas the origin of the shoulders is still not quite clear~\cite{4,5,6,7,8}. There is some evidence that the shoulders arise from intense helical (chiral) fluctuations in the vicinity of the phase transition in MnSi~\cite{8,9,10}. These fluctuations must be responsible for the first order phase transition in MnSi, symmetry of which principally allows a second order one~\cite{11,12}. The simultaneous disappearance of the shoulder and first order features at the phase transition in MnSi on decreasing temperature supports this conclusion~\cite{13}.

Insulator Cu$_2$OSeO$_3$ crystallizes in a complicated structure, which belongs to the same space group P2$_1$3 common to MnSi, and hence permits piezoelectricity~\cite{14}. On cooling below 60 K Cu$_2$OSeO$_3$ becomes magnetically ordered and demonstrates an enhanced magneto dielectric effect. The latter implies existence of the magneto electric coupling in the system. Careful structural studies of Cu$_2$OSeO$_3$ show no measurable structural distortion occurring down 10 K. This means that there is no a spontaneous lattice distortion involved in the magneto electric coupling mechanism. So a pure electronic coupling should be responsible for the observed magneto dielectric response~\cite{14}. These observations predict an incommensurate magnetic structure~\cite{15}. Finally, a helical spin structure in Cu$_2$OSeO$_3$ was found in a small angle neutron scattering experiment~\cite{16}. But despite a similar spin structure, major differences between metallic MnSi and insulating Cu$_2$OSeO$_3$ lies in their magnetic moments. The moments are itinerant in MnSi and local in Cu$_2$OSeO$_3$. Nevertheless the behavior of the heat capacity and magnetic susceptibility at the magnetic phase transitions in MnSi and Cu$_2$OSeO$_3$ appears to be very similar. Indeed the heat capacity of Cu$_2$OSeO$_3$ shows a peak and shoulder on the high temperature side of the peak at the phase transition~\cite{16} like the heat capacity of MnSi.

Actually the difference in nature of the magnetic moments between the two substances noted above reveals itself in a pressure dependence of the phase transition temperature $T_c$. $T_c$ of MnSi decreases with pressure and tends to zero, whereas $T_c$ of Cu$_2$OSeO$_3$ increases with pressure at least up to 2 GPa~\cite{20Huang}.

It would be of great interest to track an evolution of the features of the phase transitions in both materials in the hope of shedding more light on its nature and the origin of the strongly fluctuating region (shoulder). The evolution of resistivity in the vicinity of the phase transition in MnSi with pressure was analyzed in~\cite{13}.

Here we report results of a high pressure study of the heat capacity of MnSi and Cu$_2$OSeO$_3$ and the ac-magnetic susceptibility of Cu$_2$OSeO$_3$. The data obtained show different pressure but similar temperature dependences of specific features of the phase transition, therefore confirming the classical character of the strongly fluctuating region and the fluctuation nature of the first order transition in MnSi at ambient and moderate pressures.

\section{Experimental}
Single crystals of Cu$_2$OSeO$_3$ of size 0.5-1.0 mm were grown by a gas transport technique in a $610-550^\circ$C temperature gradient using a 2:1 CuO/SeO$_2$ mixture and CuCl$_2\cdot$2H$_2$O as a transport agent. High pressures were created in a small Teflon capsule filled with liquid and inserted in a miniature toroid-type clamped device~\cite{17Petrova}. Experiments with MnSi were performed with a cylinder-piston type clamped device. The specific heat C(T) at high pressure was measured by ac-calorimetry technique as described earlier~\cite{18Sidorov}. For ac-calorimetry measurements a flat zig-zag heater made of constantan wire of 12 $\mu m$ in diameter was glued to one side of a plate-like crystal ($\sim0.8\times0.8\times0.15\ mm^3$). This heater was at the same time a sensitive strain gauge and the change of its resistance under pressure allowed us to calculate the change in length and hence compressibility of the Cu$_2$OSeO$_3$ sample at room temperature as described earlier~\cite{19Tsiok}. The coil system for magnetic ac-susceptibility measurements $\chi(T)$ was placed inside the Teflon capsule. Pressure was measured by monitoring the superconducting transition temperature of Pb located near Cu$_2$OSeO$_3$ sample.

\section{Results}
\begin{figure}[htb]
\includegraphics[width=80mm]{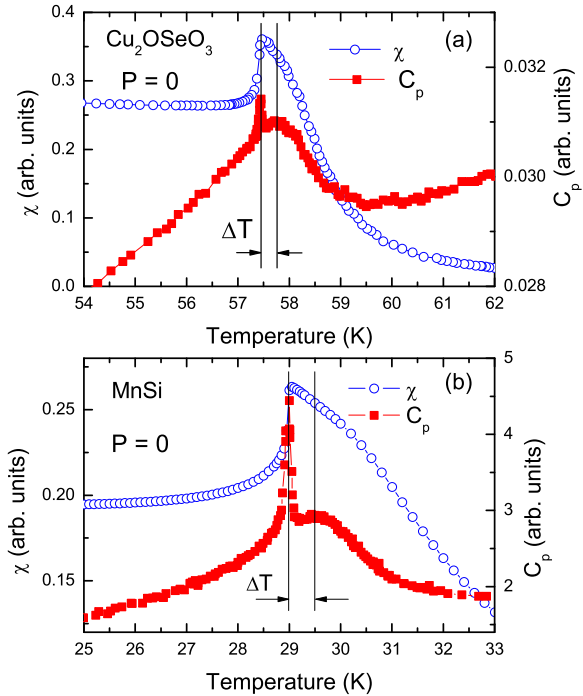}
\caption{\label{fig1} (Color online) Heat capacity $C_p$ and magnetic susceptibility $\chi$ of Cu$_2$OSeO$_3$ (a) and MnSi (b) at ambient pressure. $\Delta T$ - the temperature difference between the peaks indicated a first order phase transition and  the maxima of the anomaly, stipulated by the strong helical fluctuations.}
 \end{figure}

Fig.~\ref{fig1}a, b display the temperature dependence of the heat capacity and the magnetic susceptibility of Cu$_2$OSeO$_3$ and MnSi in the vicinity of the magnetic phase transition at ambient pressure. A striking similarity of physical properties of both substances is obvious.

\begin{figure}[htb]
\includegraphics[width=80mm]{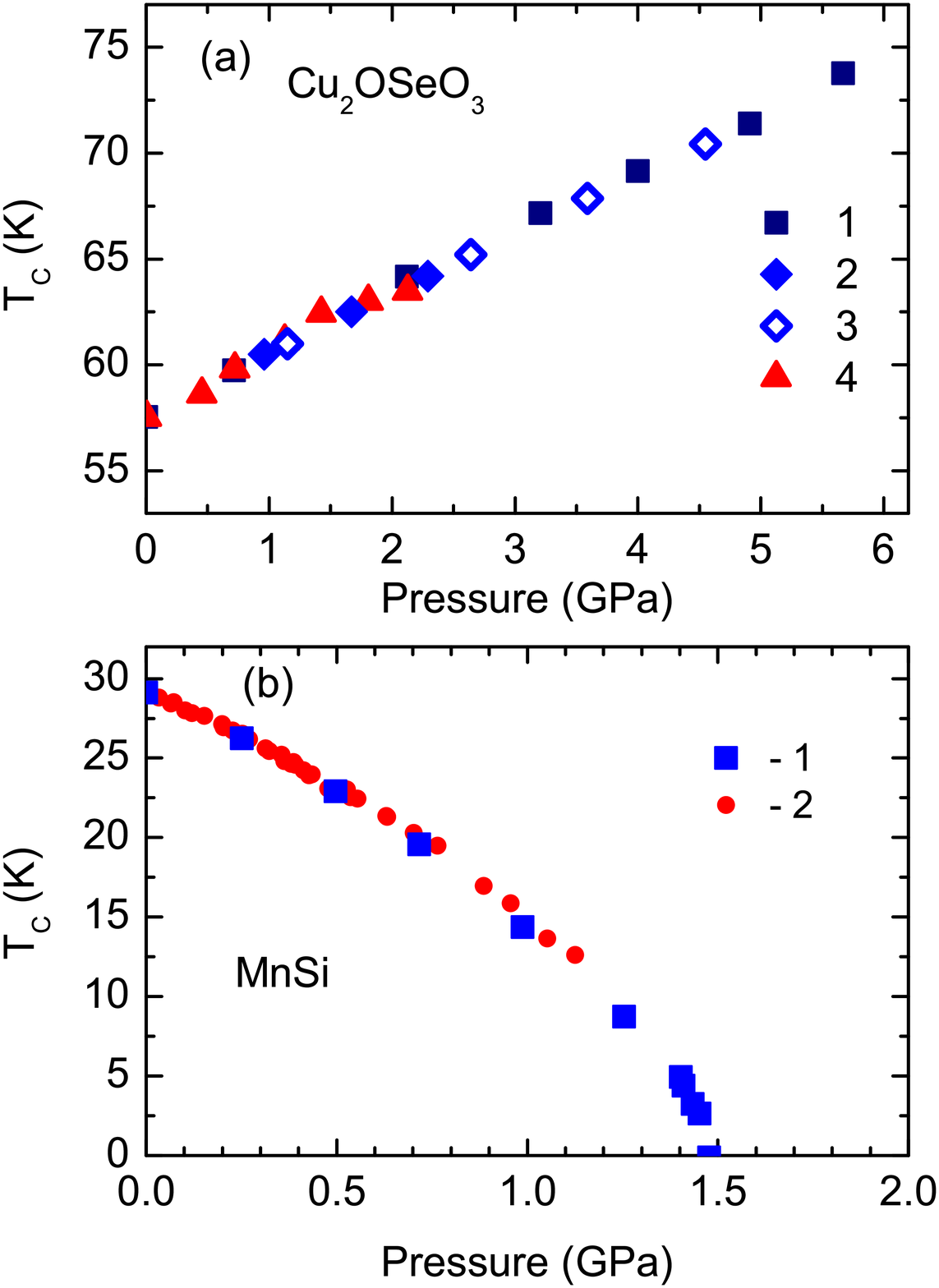}
\caption{\label{fig2} (Color online) Pressure dependence of the magnetic phase transition temperature in Cu$_2$OSeO$_3$ (a) 1- ac-susceptibility, 2- ac-calorimetry, sample 1, 3-ac-calorimetry, sample 2; 4-Ref.~\cite{20Huang} and MnSi (b) 1-Ref.~\cite{21}, 2-Ref.~\cite{22}}
 \end{figure}

\begin{figure}[htb]
\includegraphics[width=80mm]{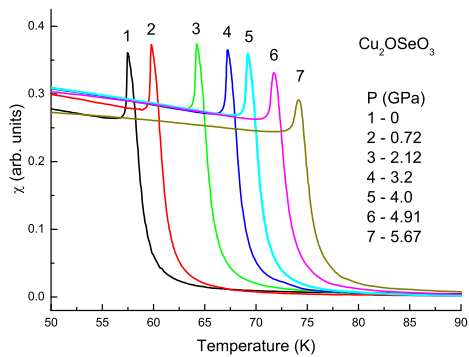}
\caption{\label{fig3} (Color online) Magnetic susceptibility $\chi$ of Cu$_2$OSeO$_3$ at different pressures.}
 \end{figure}

\begin{figure}[htb]
\includegraphics[width=80mm]{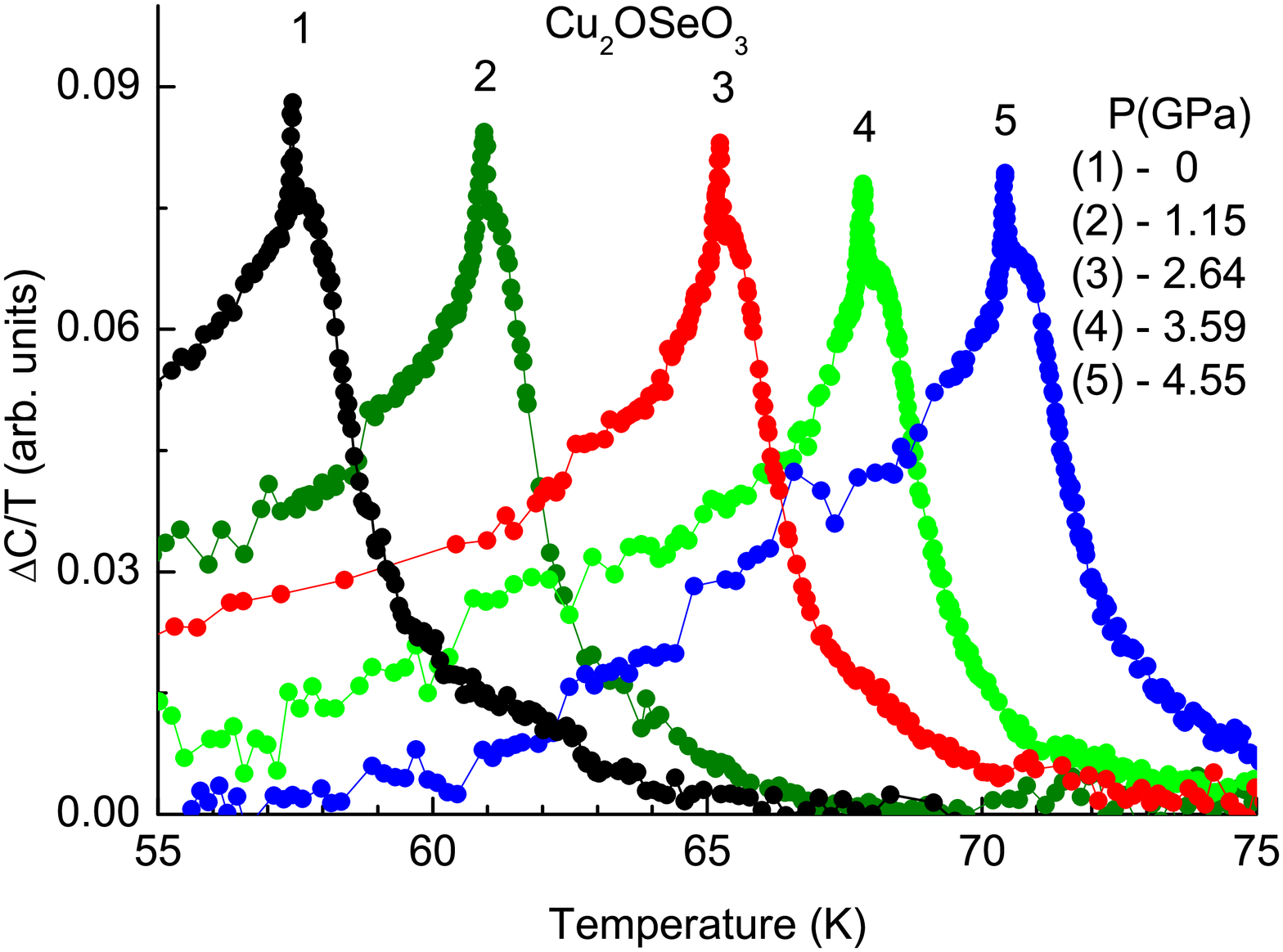}
\caption{\label{fig4} (Color online) Anomalous part of the heat capacity of Cu$_2$OSeO$_3$ at different pressures.}
 \end{figure}

\begin{figure}[htb]
\includegraphics[width=80mm]{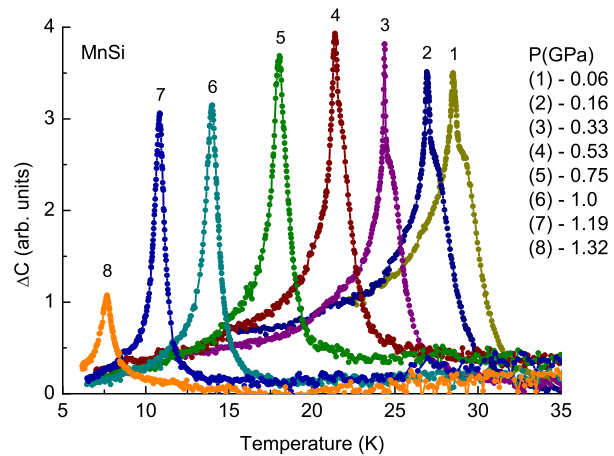}
\caption{\label{fig5} (Color online) Anomalous part of the heat capacity of MnSi at different pressures.}
 \end{figure}

Fig.~\ref{fig2}a shows the pressure dependence of the temperature of the phase transition $T_c$ of Cu$_2$OSeO$_3$, determined from magnetic susceptibility (Fig.~\ref{fig3}) and heat capacity (Fig.~\ref{fig4}) measurements. As is seen in Fig.~\ref{fig2}a, $T_c$ of Cu$_2$OSeO$_3$ increases continuously with pressure in contrast with the case of MnSi (Fig.~\ref{fig2}b). Our data on $T_c$ of Cu$_2$OSeO$_3$ are in good agreement with that obtained earlier up to 2 GPa by Huang et al.~\cite{20Huang}. Figs.~\ref{fig4} and ~\ref{fig5} show the evolution of the anomalous part of heat capacity of Cu$_2$OSeO$_3$ and MnSi with pressure and temperature.

The length of the sample of Cu$_2$OSeO$_3$ decreases linearly under pressure up to 4.5 GPa. The bulk modulus calculated from the change of the sample length is equal to $197\pm2$ GPa.


\begin{figure}[htb]
\includegraphics[width=80mm]{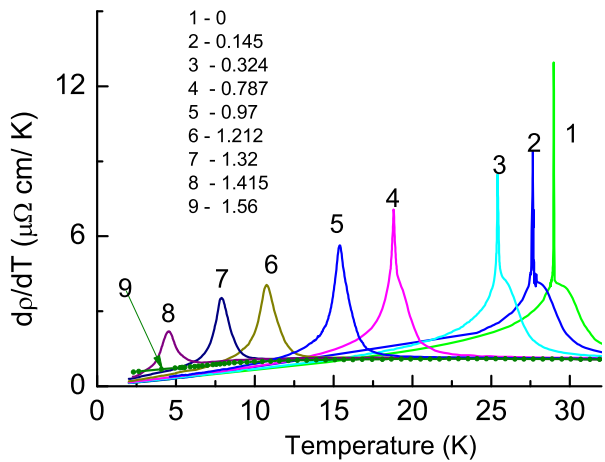}
\caption{\label{fig6} (Color online) Temperature derivatives of the resistivity $d\rho/dT$ at the phase transition in MnSi at different pressures after Ref.~\cite{13}.}
 \end{figure}
\begin{figure}[htb]
\includegraphics[width=80mm]{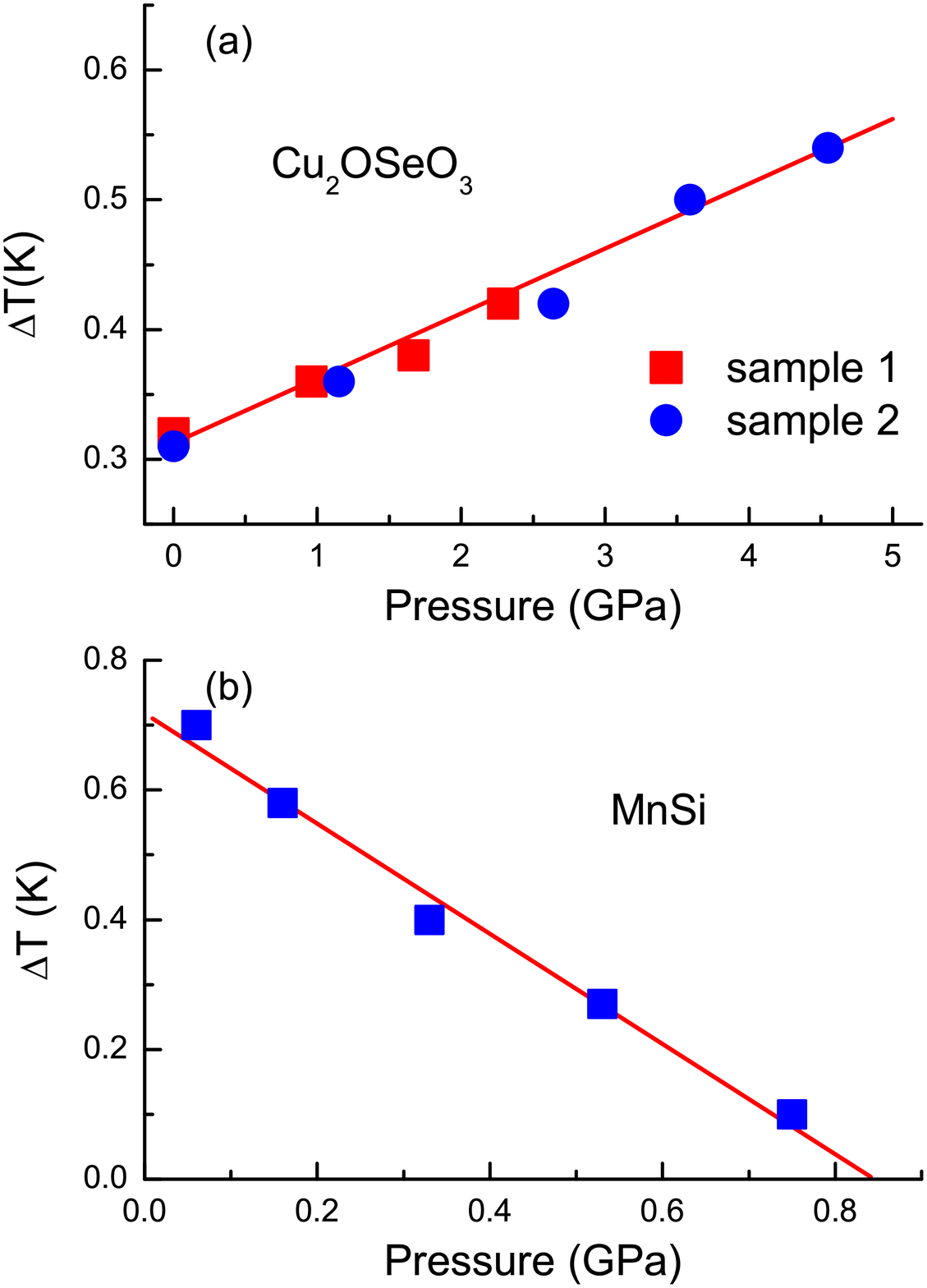}
\caption{\label{fig7} (Color online) Temperature difference $\Delta T$ between the peaks and the shoulder maxima as a function of pressure.}
 \end{figure}
 \begin{figure}[htb]
\includegraphics[width=80mm]{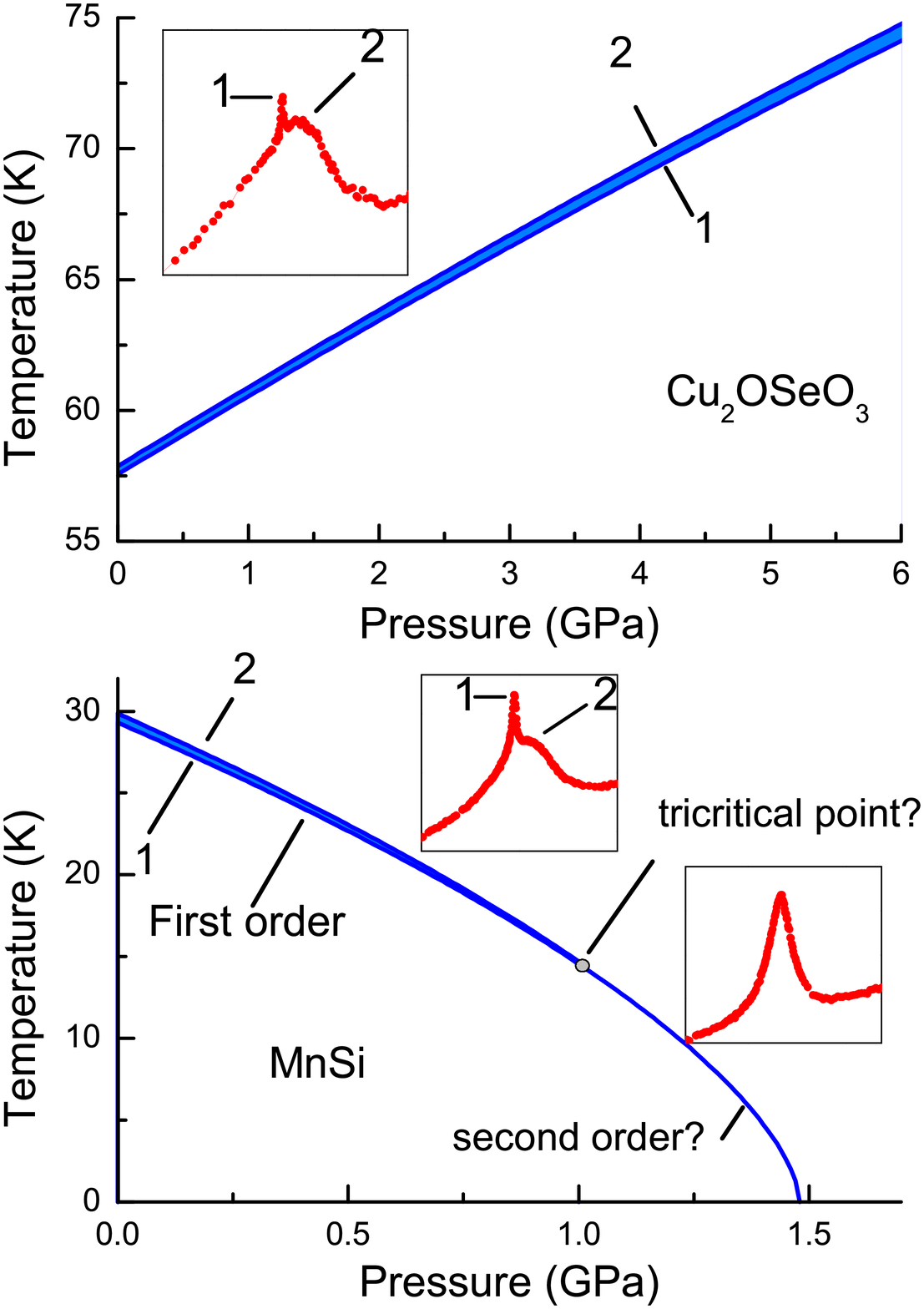}
\caption{\label{fig8} Phase diagrams of Cu$_2$OSeO$_3$ and MnSi.}
 \end{figure}

\section{Discussion}
The two substances under study belong to different classes of solids. Cu$_2$OSeO$_3$ is a covalent insulator with local magnetic moments, whereas metallic MnSi is an itinerant magnet. Normally the exchange interaction, giving rise to magnetic order in a system with local magnet moments depends on inter-particle distances in such a way that increases a phase transition temperature with pressure. This is what happens with $T_c$ of Cu$_2$OSeO$_3$ with applied pressure (see Fig.~\ref{fig2}a). In the case of itinerant magnetics an energy gain occurring at magnetic ordering arises as a result of competition between the exchange interaction and electron kinetic energy. Their changes on compression lead to a decrease of the transition temperature in itinerant magnets with pressure (see Fig.~\ref{fig2}b) Despite the mentioned difference both substances demonstrate a surprising similarity in behavior of magnetic and thermodynamic properties at the phase transition (Figs.~\ref{fig1}a, b). Probably, this implies that the longitudinal spin fluctuations typical of the weak itinerant magnets do not contribute much to the discussed properties.

Remarkably the similarity between Cu$_2$OSeO$_3$ and MnSi found at ambient pressure can be seen at high pressures as well, if one uses temperature as a variable. Indeed, analyzing Figs.~\ref{fig4} and ~\ref{fig5}, which depict the behavior of the anomalous part of heat capacity of Cu$_2$OSeO$_3$ and MnSi at different pressures and temperatures, one can see that the heat capacity peaks and shoulders clearly change at decreasing temperature. The shoulder becomes less and less prominent and completely disappeared in case of MnSi. This conclusion is obviously supported by resistivity measurements~\cite{13} (Fig.~\ref{fig6}). A variation of the temperature interval $\Delta$ between the peak and the maximum of the shoulder may serve as some sort of semi quantitative measure of this process as illustrated in Figures~\ref{fig1} and ~\ref{fig7}. Taking into account these data, phase diagrams of Cu$_2$OSeO$_3$ and MnSi were constructed in Fig.~\ref{fig8}. As can be seen, the splitting of the two characteristic features of the heat capacity curve (peak and shoulder) decreases along the phase transition line towards lower temperatures, irrespective of the different pressure dependence. Narrowing the splitting is a consequence of shrinking the heat capacity anomaly (shoulder) along with the transition temperature reduction. All this helps to identify the shoulder as a product of classical fluctuations. At the same time, in case of MnSi the sharp peaks, which classify the transition as first order cease to exist at low temperatures almost simultaneously with the disappearance of shoulder. Then with further decrease of temperature the heat capacity peaks are progressively reduced in size and width (Figs.~\ref{fig5},~\ref{fig6}). This behavior leads us to claim the existence of a tricritical point in the phase diagram of MnSi, as is shown in Fig.~\ref{fig7}b, in agreement with conclusions of Ref.~\cite{13}.

\section{Conclusion}
The heat capacity of helical magnets Cu$_2$OSeO$_3$ and MnSi has been studied at high pressures by the ac-calorimetric technique. The magnetic ac-susceptibility was measured in the vicinity of the magnetic phase transition in Cu$_2$OSeO$_3$. The helical phase transition temperature $T_c$ increases with pressure in case of Cu$_2$OSeO$_3$, which is typical of systems with local magnetic moments, whereas $T_c$ of MnSi drops on compression in accordance with standard behavior observed in itinerant magnets. The variation of two characteristic features of the heat capacity at the phase transitions of both substances (peak and shoulder) were investigated at high pressures. Despite the different nature of the magnetic moments in Cu$_2$OSeO$_3$ and MnSi these features behave in a similar way when studied as a function of temperature. So probably the longitudinal spin fluctuations can be ignored in an analysis of the phase transition in MnSi.

The shoulders shrink with decreasing temperature, which suggests that they arise from classical fluctuations. In the case of MnSi the peak and shoulder of the heat capacity disappear simultaneously probably signifying the existence of a tricritical point and confirming the fluctuation nature of the first order phase transition in Cu$_2$OSeO$_3$ and MnSi.

\section{Acknowledgements}
This work was supported by the Russian Foundation for Basic Research (grant 12-02-00376-a), Program of the Physics Department of RAS on Strongly Correlated Electron Systems and Program of the Presidium of RAS on Strongly Compressed Matter.

\end{document}